\documentclass[prl,twocolumn,superscriptaddress,nopacs]{revtex4}
\usepackage{times}
\usepackage{amsmath}
\usepackage{graphicx}
\usepackage{pst-all}

\begin{document}

\title{Decoherence of Anyonic Charge in Interferometry
Measurements}
\author{Parsa Bonderson}
\affiliation{California Institute of Technology, Pasadena, CA 91125}
\author{Kirill Shtengel}
\affiliation{Department of Physics and Astronomy, University of California, Riverside, CA 92521}
\affiliation{California Institute of Technology, Pasadena, CA 91125}
\author{J. K. Slingerland}
\affiliation{Microsoft Research, Station Q, CNSI Building, University of California, Santa Barbara, CA 93106}
\affiliation{Department of Physics and Astronomy, University of California, Riverside, CA 92521}
\affiliation{California Institute of Technology, Pasadena, CA 91125}
\date{\today}

\begin{abstract}
We examine interferometric measurements of the topological charge of
(non-Abelian) anyons. The target's topological charge is measured from its
effect on the interference of probe particles sent through the interferometer.
We find that superpositions of distinct anyonic charges $a$ and $a^{\prime}$ in
the target decohere (exponentially in the number of probes particles used) when
the probes have nontrivial monodromy with the charges that may be fused with $a$
to give $a^{\prime}$.
\end{abstract}


\maketitle


Quantum physics in two spatial dimensions allows for the existence of
particles which are neither bosons nor fermions. Instead, the exchange
interactions of such ``anyons'' are described by representations of the
braid group \cite{Leinaas77,Wilczek82a,Wilczek82b}, which may even be
non-Abelian \cite{Goldin85,Froehlich90}. Recently, there has
been a resurgence of interest in anyons, due to increased experimental
capabilities in systems believed to harbor them, and also their potential
application to topologically protected quantum computation
\cite{Kitaev03,Preskill98,Freedman02a}. In this quantum
computing scheme, qubits are encoded in non-localized, topological
charges carried by clusters of non-Abelian anyons. Topological charges decouple
from local probes, affording them protection from decoherence. However, this
also makes their measurement, which is vital for qubit readout, more difficult,
typically requiring interferometry. The most promising
candidate system for discovering non-Abelian statistics is the fractional
quantum Hall (FQH) state observed at filling fraction $\nu =5/2$
\cite{Willett87,Pan99}, which is widely expected to be described by the
Moore-Read state \cite{Moore91,Nayak96c}. Interference experiments, similar to
that proposed \cite{Chamon97} and only
recently implemented \cite{Camino05a,Camino05b} for \emph{Abelian} FQH
states, may soon verify the braiding statistics of the $\nu =5/2$ state \cite
{Fradkin98,DasSarma05,Stern06a,Bonderson06a}. The analyses in these
treatments assume the target particle to be in an eigenstate of topological
charge. We show that, when this is not the case, the density matrix of the
target particle is diagonalized in the charge basis during the experiment
if a simple criterion on the braiding of source and target particles is
satisfied: superpositions of distinct anyonic charges $a$ and $a^{\prime }$
decohere as long as the probe particles have nontrivial monodromy with the
charge differences between $a$ and $a^{\prime }$, that is, with the charges
that fuse with $a$ to give $a^{\prime }$.

\begin{figure}[hbt]
\includegraphics[width=2in]{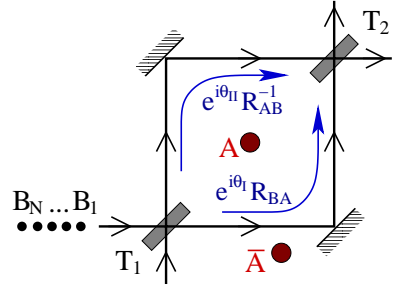}
\caption{A Mach-Zehnder interferometer containing the target anyon(s) $A$, to be
probed by the anyons $B_k$. (Detectors not shown.)}
\label{fig:interferometer}
\end{figure}

We consider a Mach-Zehnder type interferometer
(see Fig. \ref{fig:interferometer}), though the same methods may be applied
to other types, with similar conclusions. A target ``particle'' $A$ carrying a
superposition of anyonic charges \footnote{Since localized charges cannot be
superimposed, when we refer to ``a particle having a superposition of charges,''
we really mean several (quasi-)particles, treated collectively.} is located in
the region between the two paths of the
interferometer. A beam of probe particles $B_{k}$, $k=1,\ldots ,N$ may be
sent into two possible input channels, is passed through a beam splitter $%
T_{1}$, reflected by mirrors around the central region, passed through a
second beam splitter $T_{2}$, and finally detected at two possible output
channels. The state acquires a phase $e^{i\theta _{\text{I}}}$ or $e^{i\theta
_{\text{II}}}$ when a probe particle passes through the bottom or top path
around
the central region (this may come from background flux, path length
differences, phase shifters, etc.) and a separate, independent contribution
strictly from the braiding of the probe and target particles, which, for
non-Abelian anyons, will be more complicated than a mere phase. If the
phases $e^{i\theta _{\text{I}}}$ and $e^{i\theta _{\text{II}}}$ are fixed, or
closely
monitored, this provides a non-demolitional measurement of
the anyonic charge of $A$ \footnote{\emph{Decoherence} in the charge basis
is actually independent of these phases, as they drop out of the
density matrix for $A$.}. This admittedly idealized setup is similar to
one experimentally realized for quantum Hall systems \cite{Ji03}, the
primary difference being that the number of quasiparticle excitations in the
central interferometry region is not fixed in that experiment. While
unsuitable for measuring a target charge, this situation may still be used to
detect the presence of non-Abelian statistics \cite{Feldman06}.

The experiment we describe was also considered in the paper
\cite{Overbosch01}, where it was referred to as the ``many-to-one''
experiment. In that paper, the authors use a quantum group inspired approach,
where individual particles are assumed to have internal Hilbert spaces, and
they study what happens to the internal state of the target particle. In our
descriptions of the systems examined, we use the theory of general anyon
models (unitary braided tensor categories), which does not ascribe individual
particles internal degrees of freedom. Instead, the relevant observables are the
overall anyonic charges of groups of particles (our main result will be stated
in terms of the density matrix of an anyon pair $A$--$\overline{A}$).
This is the situation relevant to the topological systems (e.g. FQH states)
that we have in mind. We also remove some constraints imposed in
\cite{Overbosch01}, specifically, that the probe particles are all identical
and have trivial self-braiding.

Let us recall some information about anyon models (see e.g.
\cite{Preskill-lectures,Kitaev06a} for additional details). States in these
models may be represented by superpositions
of oriented worldline diagrams that give a history of splitting and fusion
of particles carrying an anyonic charge. Each allowed fusion/splitting vertex
is associated with a (possibly multi-dimensional) vector space containing
normalized bra/ket vectors%
\begin{eqnarray}
\left( d_{c} / d_{a}d_{b} \right) ^{1/4}
\pspicture[0.4](0,0)(1.4,-1)
  \psset{linewidth=0.9pt,linecolor=black,arrowscale=1.5,arrowinset=0.15}
  \psline{-<}(0.7,0)(0.7,-0.35)
  \psline(0.7,0)(0.7,-0.55)
  \psline(0.7,-0.55) (0.25,-1)
  \psline{-<}(0.7,-0.55)(0.35,-0.9)
  \psline(0.7,-0.55) (1.15,-1)	
  \psline{-<}(0.7,-0.55)(1.05,-0.9)
  \rput[tl]{0}(0.4,0){$c$}
  \rput[br]{0}(1.4,-0.95){$b$}
  \rput[bl]{0}(0,-0.95){$a$}
  \rput[bl]{0}(0.85,-0.5){$\mu$}
  \endpspicture
&=&\left\langle a,b;c,\mu \right| \in
V_{ab}^{c}
\label{eq:bra}\\
\left( d_{c} / d_{a}d_{b}\right) ^{1/4}
\pspicture[0.4](0,0)(1.4,1)
  \psset{linewidth=0.9pt,linecolor=black,arrowscale=1.5,arrowinset=0.15}
  \psline{->}(0.7,0)(0.7,0.45)
  \psline(0.7,0)(0.7,0.55)
  \psline(0.7,0.55) (0.25,1)
  \psline{->}(0.7,0.55)(0.3,0.95)
  \psline(0.7,0.55) (1.15,1)	
  \psline{->}(0.7,0.55)(1.1,0.95)
  \rput[bl]{0}(0.4,0){$c$}
  \rput[br]{0}(1.4,0.8){$b$}
  \rput[bl]{0}(0,0.8){$a$}
  \rput[bl]{0}(0.85,0.35){$\mu$}
  \endpspicture
&=&\left| a,b;c,\mu \right\rangle \in
V_{c}^{ab},
\label{eq:ket}
\end{eqnarray}%
where $\mu $ labels the basis states of the splitting space $V_{c}^{ab}$ of
the charges $a$ and $b$ from charge $c$ and the number $d_{a}\ge 1$ is the
quantum dimension of $a$. The factors of $\left( d_{c} / d_{a}d_{b}\right)
^{1/4}$ are necessary for isotopy invariance, i.e. so the meaning of the
diagrams is not changed by continuous deformation. The vacuum is labeled $1$,
and has $d_{1}=1$. Since $\dim V_{c}^{ab}=1$ when any of $a,b,c$ equals $1$,
the basis label in this case is redundant, and will be dropped. In fact, the meaning of
diagrams is invariant under addition/removal of vacuum lines, so we may
drop them and smooth out their vertices. The charge conjugate, or antiparticle,
of $a$ is denoted $\overline{a}$, and may also be denoted by reversing the arrow
on a line labeled by $a$. Diagrams with multiple vertices correspond to tensor
products of vertex spaces. Density matrices may be represented by diagrams
with the same numbers of lines emerging at the top and bottom (being
combinations of kets and bras). Conjugation of states and operators
corresponds to reflecting their diagrams through the horizontal plane while
reversing orientations [e.g. Eqs.~(\ref{eq:bra}),(\ref{eq:ket})]. One may
diagrammatically trace out a charge that enters and exits a diagram at the
same spatial position by connecting the lines at these positions with an arc
that does not interfere with the rest of the diagram (giving zero if the
charges do not match). This is actually the \emph{quantum} trace, which equals
the ordinary trace with each sector of overall charge $f$ multiplied by $d_f$.
Here are some important diagrammatic relations:%
\begin{equation}
\pspicture[0.4](0,-0.4)(1.1,1)
  \psset{linewidth=0.9pt,linecolor=black,arrowscale=1.5,arrowinset=0.15}
  \psline(0.3,-0.45)(0.3,1)
  \psline{->}(0.3,-0.45)(0.3,-0.05)
  \psline{->}(0.3,0.5)(0.3,0.85)
  \psline(0.8,-0.45)(0.8,1)
  \psline{->}(0.8,-0.45)(0.8,-0.05)
  \psline{->}(0.8,0)(0.8,0.85)
  \psline(0.8,0.05)(0.3,0.45)
  \psline{->}(0.8,0.05)(0.45,0.33)
  \rput[bl]{0}(0.48,0.4){$e$}
  \rput[br]{0}(1.05,0.8){$b$}
  \rput[bl]{0}(0,0.8){$a$}
  \rput[bl]{0}(0.02,0.35){$\alpha$}
  \rput[br]{0}(1.1,-0.4){$c$}
  \rput[bl]{0}(-0.05,-0.4){$d$}
  \rput[bl]{0}(0.85,-0.1){$\beta$}
  \endpspicture
=\sum\limits_{f,\mu
,\nu }\left[ F_{d,c}^{a,b}\right] _{\left( e,\alpha ,\beta \right) ,\left(
f,\mu ,\nu \right)}
\pspicture[0.4](0,-0.4)(1.4,1)
  \psset{linewidth=0.9pt,linecolor=black,arrowscale=1.5,arrowinset=0.15}
  \psline{->}(0.7,0)(0.7,0.45)
  \psline(0.7,0)(0.7,0.55)
  \psline(0.7,0.55) (0.25,1)
  \psline{->}(0.7,0.55)(0.3,0.95)
  \psline(0.7,0.55) (1.15,1)	
  \psline{->}(0.7,0.55)(1.1,0.95)
  \rput[bl]{0}(0.38,0.2){$f$}
  \rput[br]{0}(1.4,0.8){$b$}
  \rput[bl]{0}(0,0.8){$a$}
  \rput[bl]{0}(0.85,0.35){$\mu$}
  \psline(0.7,0) (0.25,-0.45)
  \psline{-<}(0.7,0)(0.35,-0.35)
  \psline(0.7,0) (1.15,-0.45)	
  \psline{-<}(0.7,0)(1.05,-0.35)
  \rput[br]{0}(1.4,-0.4){$c$}
  \rput[bl]{0}(0,-0.4){$d$}
  \rput[bl]{0}(0.85,-0.05){$\nu$}
  \endpspicture,
\end{equation}
\begin{equation}
R_{ab}=
\pspicture[0.4](0,0)(1.2,1)
  \psset{linewidth=0.9pt,linecolor=black,arrowscale=1.5,arrowinset=0.15}
  \psline(0.96,0.05)(0.2,1)
  \psline{->}(0.96,0.05)(0.28,0.9)
  \psline(0.24,0.05)(1,1)
  \psline[border=2pt]{->}(0.24,0.05)(0.92,0.9)
  \rput[bl]{0}(0,0){$a$}
  \rput[br]{0}(1.2,0){$b$}
  \endpspicture
,\qquad
R_{ab}^{\dag}= R_{ab}^{-1}=
\pspicture[0.4](0,0)(1.2,1)
  \psset{linewidth=0.9pt,linecolor=black,arrowscale=1.5,arrowinset=0.15}
  \psline{->}(0.24,0.05)(0.92,0.9)
  \psline(0.24,0.05)(1,1)
  \psline(0.96,0.05)(0.2,1)
  \psline[border=2pt]{->}(0.96,0.05)(0.28,0.9)
  \rput[bl]{0}(0,0){$b$}
  \rput[br]{0}(1.2,0){$a$}
  \endpspicture
,
\end{equation}
\begin{equation}
S_{ab}=\frac{1}{D}
\pspicture[0.5](2.4,1.3)
  \psarc[linewidth=0.9pt,linecolor=black,arrows=<-,arrowscale=1.5,
arrowinset=0.15] (1.6,0.7){0.5}{165}{363}
  \psarc[linewidth=0.9pt,linecolor=black] (0.9,0.7){0.5}{0}{180}
  \psarc[linewidth=0.9pt,linecolor=black,border=3pt,arrows=->,arrowscale=1.5,
arrowinset=0.15] (0.9,0.7){0.5}{180}{375}
  \psarc[linewidth=0.9pt,linecolor=black,border=3pt] (1.6,0.7){0.5}{0}{160}
  \psarc[linewidth=0.9pt,linecolor=black] (1.6,0.7){0.5}{155}{170}
  \rput[bl]{0}(0.15,0.3){$a$}
  \rput[bl]{0}(2.15,0.3){$b$}
  \endpspicture
, \qquad
\pspicture[0.4](1,1.3)
  \psset{linewidth=0.9pt,linecolor=black,arrowscale=1.5,arrowinset=0.15}
  \psline(0.4,0)(0.4,1.2)
  \psarc[linewidth=0.9pt,linecolor=black] (0.4,0.5){0.3}{3}{180}
\psarc[linewidth=0.9pt,linecolor=black,border=2.5pt,arrows=->,arrowscale=1.4,
arrowinset=0.15] (0.4,0.5){0.3}{180}{375}
\psline[linewidth=0.9pt,linecolor=black,border=2.5pt,arrows=->,arrowscale=1.5,
arrowinset=0.15](0.4,0.5)(0.4,1.1)
  \rput[bl]{0}(0.8,0.4){$a$}
  \rput[tl]{0}(0.5,1.2){$b$}
  \endpspicture
=\frac{S_{ab}}{S_{1b}}
\pspicture[0.4](0.3,0)(0.6,1.3)
  \psset{linewidth=0.9pt,linecolor=black,arrowscale=1.5,arrowinset=0.15}
  \psline(0.4,0)(0.4,1.2)
\psline[linewidth=0.9pt,linecolor=black,arrows=->,arrowscale=1.5,
arrowinset=0.15](0.4,0.5)(0.4,0.9)
  \rput[tl]{0}(0.5,1.0){$b$}
  \endpspicture
\label{eq:loopaway}
\end{equation}%
where $d_a=DS_{1a}$ is the value of an unknotted loop carrying charge $a$, and
$D=\sqrt{\sum_{a}d_{a}^{2}}$ is the total quantum dimension. Another useful
quantity, especially for interference experiments \cite{Bonderson06b},
is the monodromy matrix element $M_{ab}=\frac{S_{ab}S_{11}}{S_{1a}S_{1b}}$.
It has the property $\left| M_{ab}\right| \leq 1$, with $M_{ab}=1$ corresponding
to trivial monodromy, i.e. the state is unchanged by taking the
charges $a$ and $b$ all the way around each other.

Using this formalism, it is important to keep track of all particles involved in
a process. We invoke the physical assumption that the particles $A$ and all
$B_{k}$ are initially unentangled. This means there are no non-trivial charge
lines connecting them, and to achieve this, they must each be created separately
from vacuum, with their own antiparticles \footnote{These
``particle-antiparticle pairs'' may really be multiple pair-created particles
that are made to interact amongst each other as needed and then split into two
groups that are henceforth treated collectively.}. We write the initial state of
the $A$--$\overline{A}$ system as%
\begin{equation}
\left| \Psi _{0}\right\rangle =\sum\limits_{a}A_{a}\left| a,\overline{a}%
;1\right\rangle
\end{equation}%
and that of each $B_{k}$--$\overline{B}_{k}$ system as%
\begin{equation}
\label{eq:stateBk}
\left| \varphi _{k}\right\rangle =\sum\limits_{b,s}B_{b,s}^{\left( k\right)
}\left| \overline{b},b;1;s\right\rangle
\end{equation}%
where $s=\rightarrow ,\uparrow $ indicates in which direction the probe
particle is traveling. The probes' antiparticles, $\overline{B}_{k}$,
will be taken off to the left and do not participate in the interferometry.
The location of the target's antiparticle $\overline{A}$ with respect to
the interferometer is important and we will let it be located below the
central region, as in Fig. \ref{fig:interferometer}.

Utilizing the two-component vector notation $
\left(
\begin{smallmatrix}
1 \\
0
\end{smallmatrix}
\right) =\left| \rightarrow \right\rangle$, $\left(
\begin{smallmatrix}
0 \\
1
\end{smallmatrix}
\right) =\left| \uparrow \right\rangle$, the two beam splitters, which (along
with the mirrors) are assumed to be lossless, are represented by the unitary
operators
$T_{j}= \left[ \begin{smallmatrix}
t_{j} & r_{j}^{\ast } \\
r_{j} & -t_{j}^{\ast }
\end{smallmatrix} \right]$, with $\left| t_{j}\right| ^{2}+\left|
r_{j}\right| ^{2}=1$.
The evolution operator for passing the probe particle $B_{k}$ through the
interferometer is%
\begin{eqnarray}
U_{k} &=&T_{2}\Sigma _{k}T_{1} \\
\Sigma _{k} &=&\left[
\begin{array}{cc}
0 & e^{i\theta _{\text{II}}}R_{A,B_{k}}^{-1} \\
e^{i\theta _{\text{I}}}R_{B_{k},A} & 0
\end{array}%
\right].
\end{eqnarray}%
Diagrammatically, this takes the form%
\begin{multline}
\label{eq:Udiag}
\pspicture[0.4](-0.2,0)(1,1)
\rput[tl](0,0){$B_{k}$}
\rput[tl](0,1.2){$A$}
\rput[tl](0.85,1.2){$B_{k}$}
\rput[tl](0.85,0){$A$}
 \psline[linewidth=0.9pt](0.4,0)(0.4,0.24)
 \psline[linewidth=0.9pt](0.78,0)(0.78,0.24)
 \psline[linewidth=0.9pt](0.4,0.78)(0.4,1)
 \psline[linewidth=0.9pt](0.78,0.78)(0.78,1)
\rput[bl](0.25,0.24){\psframebox{$U_k$}}
  \endpspicture
=e^{i\theta _{\text{I}}}\left[
\begin{array}{cc}
t_{1}r_{2}^{\ast } & r_{1}^{\ast }r_{2}^{\ast } \\
-t_{1}t_{2}^{\ast } & -r_{1}^{\ast }t_{2}^{\ast }
\end{array}
\right]
\pspicture[0.35](-0.2,0)(1.25,1)
  \psset{linewidth=0.9pt,linecolor=black,arrowscale=1.5,arrowinset=0.15}
  \psline(0.92,0.1)(0.2,1)
  \psline{->}(0.92,0.1)(0.28,0.9)
  \psline(0.28,0.1)(1,1)
  \psline[border=2pt]{->}(0.28,0.1)(0.92,0.9)
  \rput[tl]{0}(-0.2,0.2){$B_k$}
  \rput[tr]{0}(1.25,0.2){$A$}
  \endpspicture
 \\
+e^{i\theta _{\text{II}}}\left[
\begin{array}{cc}
r_{1}t_{2} & -t_{1}^{\ast }t_{2} \\
r_{1}r_{2} & -t_{1}^{\ast }r_{2}
\end{array}
\right]
\pspicture[0.4](-0.2,0)(1.25,1)
  \psset{linewidth=0.9pt,linecolor=black,arrowscale=1.5,arrowinset=0.15}
  \psline{->}(0.28,0.1)(0.92,0.9)
  \psline(0.28,0.1)(1,1)
  \psline(0.92,0.1)(0.2,1)
  \psline[border=2pt]{->}(0.92,0.1)(0.28,0.9)
  \rput[tl]{0}(-0.2,0.2){$B_k$}
  \rput[tr]{0}(1.25,0.2){$A$}
  \endpspicture
  .
\end{multline}%
Keeping track of antiparticles, we need $V_{k}=R_{\overline{A},B_{k}}^{-1}$ for
braiding the probe particles with $\overline{A}$ \footnote{If $\overline{A}$ is
located above, rather than below, the central region of the interferometer, we
would instead use $V_{k}=R_{B_{k},\overline{A}}$. This essentially interchanges
$r_{1}$ with $t_{1}$ and conjugates $M_{be}$ in the result,
Eq. (\ref{eq:rhoAN}). If however, $\overline{A}$ is placed between the
two output legs, the situation is complicated by the resulting
$V_{k}= \left[ \protect\begin{smallmatrix}
R_{B_{k},\overline{A}} & 0 \\
0 & R_{\overline{A},B_{k}}^{-1}
\protect\end{smallmatrix} \right]$,
which makes evaluation more difficult, and gives a different limiting
behavior. If $\overline{A}$ is situated in the central region (with $A$),
there will, of course, be no effect.}, and, adding in each successive
$\left| \varphi _{k}\right\rangle$ from the left, we also use the operators%
\begin{equation}
W_{k}=R_{\overline{B}_{k},\overline{B}_{k-1}}R_{B_{k},\overline{B}
_{k-1}}\ldots R_{\overline{B}_{k},\overline{B}_{1}}R_{B_{k},\overline{B}_{1}}
\end{equation}%
(and $W_1=1$), which move the
$\overline{B}_{k}B_{k}$ pair from the left to the center of the configuration
$\overline{B}_{1}\ldots \overline{B}_{k-1}B_{k-1}\ldots B_{1}$.
This may be viewed either as spatial sorting after creation, or, as shown
suggestively in Eq.~(\ref{eq:state}),
as the temporal condition that each $\overline{B}B$\ pair is utilized
before creating the next.

The state of the combined system after $N$ probe
particles have passed through the interferometer (but have not yet been
detected) may now be defined iteratively as%
\begin{equation}
\left| \Psi _{N}\right\rangle =V_{N}U_{N}W_{N}\left| \varphi
_{N}\right\rangle \otimes \left| \Psi _{N-1}\right\rangle
.
\end{equation}

Focusing on the $A$--$\overline{A}$ system, the reduced density matrix,
$\rho _{N}^{A}=Tr_{B^{\otimes N}}\left[ \left| \Psi_{N}\right\rangle \left\langle \Psi _{N}\right| \right] $,
is obtained by tracing over the $B_{k}$ and $\overline{B}_{k}$ particles.
This may be interpreted as ignoring the detection results. Given the placement
of $\overline{A}$, one sees that this averaging over detector measurements makes
the second beam splitter irrelevant. If we kept track of the measurement
outcomes $s_k$, we would project with
$ \left| s_{k}\right\rangle \left\langle s_{k}\right| $ after the
$k^{th}$ probe particle. In $\left| \Psi_{N}\right\rangle$, we did not include
braidings between the $B_{k}$, but they may be added
without changing the results, as they drop out of $\rho _{N}^{A} $
\footnote{\emph{Superpositions} of braiding may however change these results.}.

We will first assume that the probe particles all have the same,
definite anyonic charge $b$ and enter through the horizontal leg, so that
$\left| \varphi _{k}\right\rangle =\left| \overline{b},b;1;\rightarrow
\right\rangle $ for all $k$, and then later return to the general case.
This results in the state%
\begin{equation}
\left| \Psi_N \right. \rangle =
\sum_{a} A_a \frac{1}{\sqrt{d_a d_b^N}}
\pspicture[0.45](-0.1,0.175)(3.1,2)
\scriptsize
\psset{linewidth=0.9pt,arrowscale=1.3,arrowinset=0.15}
\rput[bl](0,2){\rnode{B1}{$\overline{b}$}}
\rput[bl](0.13,2){\rnode{B2}{\tiny$\ldots$}}
\rput[bl](0.49,2){\rnode{Bn}{$\overline{b}$}}
\rput[bl](0.9,2){\rnode{A1}{$a$}}
\rput[bl](2.1,2){\rnode{A2}{$\overline{a}$}}
\rput[bl](2.5,2){\rnode{B11}{$b$}}
\rput[bl](2.63,2){\rnode{B12}{\tiny$\ldots$}}
\rput[bl](2.99,2){\rnode{B1n}{$b$}}
\rput[bl](0.9,1.05){\rnode{Un}{\psframebox{$\!U_N\!\!$}}}
\rput[bl](1.25,0.45){\rnode{U1}{\psframebox{$U_1\!$}}}
\pnode(0.95,0.175){O1}
\pnode(0.95,0.75){On}
\pnode(1.8,0.175){Oa}
\pnode(1.25,1.07){Oy}
\pnode(1.45,1.0){Ox}
\nccurve[nodesepA=2pt,angleA=-90,angleB=135]{B1}{O1}
\nccurve[nodesepA=2pt,angleA=-90,angleB=135]{Bn}{On}
\nccurve[angleA=-115,angleB=45]{U1}{O1}
\nccurve[angleA=-105,angleB=45]{Un}{On}
\nccurve[angleA=108,angleB=-72,linestyle=dotted,linewidth=1pt,dotsep=1.4pt]
{U1} {Un}
\nccurve[nodesepA=2pt,angleA=-105,angleB=75]{B1n}{U1}
\nccurve[nodesepA=2pt,angleA=-105,angleB=55]{B11}{Un}
\nccurve[nodesepA=2pt,angleA=-95,angleB=65,border=2pt]{A2}{Oa}
\nccurve[angleA=-75,angleB=125]{U1}{Oa}
\nccurve[nodesepA=2pt,angleA=-85,angleB=105]{A1}{Un}
\psline{->}(0.0705,1.775)(0.065,1.825)
\psline{->}(0.556,1.775)(0.552,1.825)
\psline{->}(1.026,1.775)(1.008,1.825)
\psline{->}(2.138,1.675)(2.1425,1.725)
\psline{->}(1.75,1.706)(1.8,1.7175)
\psline{->}(2.825,1.675)(2.89,1.725)
\endpspicture
\label{eq:state}
\end{equation}%
\vspace{0.5cm}
(with directional indices left implicit).

We first consider the case $N=1$. Tracing out the $b$ and $\overline{b}$ lines
of $\left| \Psi_{1}\right\rangle \left\langle \Psi _{1}\right|$, and using
Eq.~(\ref{eq:Udiag}), one finds that terms cancel to give%
\begin{multline}
\rho_1^A =
\sum_{a,a'} \frac{A_a A_{a'}^{\ast}}{\sqrt{d_a d_{a'}} d_b }
\\
\times
\left[
\left|r_1 \right|^2\,
\pspicture[0.45](-0.3,-0.48)(1.5,1.26)
\scriptsize
\psset{linewidth=0.9pt,arrowscale=1.3,arrowinset=0.15}
\pnode(0,0){LB}
\pnode(0,0.96){LT}
\pnode(1.2,-0.06){RB}
\pnode(1.2,1.02){RT}
\pnode(0.12,0.6){OB1}
\pnode(0.12,0.36){OB2}
\pnode(0.36,0.78){MR}
\pnode(0.34,0.77){MR2}
\pnode(0.36,0.18){ML}
\pnode(0.75,0.6){OA1}
\pnode(0.75,0.36){OA2}
\nccurve[angleA=135,angleB=-80]{->}{OB1}{LT}
\nccurve[angleA=80,angleB=-135]{LB}{OB2}
\nccurve[angleA=80,angleB=-135]{>-}{LB}{OB2}
\pscurve(0.012,0.9)(-0.12,1.14)(-0.3,0.48)(-0.12,-0.18)(0.012,0.06)
\nccurve[angleA=55,angleB=-148]{->}{OB1}{MR}
\nccurve[angleA=25,angleB=-120]{MR2}{RT}
\pscurve(1.2,1.02)(1.32,1.14)(1.5,0.48)(1.32,-0.18)(1.2,-0.06)
\nccurve[angleA=155,angleB=-55]{ML}{OB2}
\nccurve[angleA=155,angleB=-55]{>-}{ML}{OB2}
\nccurve[angleA=120,angleB=-25]{RB}{ML}
\rput[bl](-0.18,0.57){$\overline{b}$}
\rput[bl](0.24,0.84){$b$}
\rput[bl](-0.18,0.09){$\overline{b}$}
\rput[bl](0.24,-0.09){$b$}
\rput[bl](0.45,1.14){\rnode{A1}{$a$}}
\rput[bl](0.93,1.14){\rnode{A2}{$\overline{a}$}}
\ncarc[nodesepA=2pt,linewidth=0,border=2pt]{A2}{OA1}
\ncarc[nodesepB=1.5pt,border=2pt]{OA1}{A1}
\ncarc[nodesepA=1.5pt,border=2.5pt]{A2}{OA1}
\ncarc[nodesepB=2pt]{->}{OA1}{A1}
\ncarc[nodesepA=2pt]{<-}{A2}{OA1}
\rput[bl](0.45,-0.42){\rnode{A3}{$a'$}}
\rput[bl](0.93,-0.42){\rnode{A4}{$\overline{a'}$}}
\ncarc[nodesepA=2pt,border=2pt]{A3}{OA2}
\ncarc[nodesepB=2pt,border=2pt]{OA2}{A4}
\pnode(0.61,-0.06){XX}
\pnode(0.97,-0.06){YY}
\ncarc{-<}{OA2}{YY}
\ncarc{>-}{XX}{OA2}
\endpspicture
+\left|t_1 \right|^2\,
\pspicture[0.45](-0.3,-0.48)(1.5,1.26)
\scriptsize
\psset{linewidth=0.9pt,arrowscale=1.3,arrowinset=0.15}
\pnode(0,0){LB}
\pnode(0,0.96){LT}
\pnode(1.2,-0.06){RB}
\pnode(1.2,1.02){RT}
\pnode(0.12,0.6){OB1}
\pnode(0.12,0.36){OB2}
\pnode(0.36,0.78){MR}
\pnode(0.34,0.77){MR2}
\pnode(0.36,0.18){ML}
\pnode(0.75,0.6){OA1}
\pnode(0.75,0.36){OA2}
\nccurve[angleA=135,angleB=-80]{->}{OB1}{LT}
\nccurve[angleA=80,angleB=-135]{LB}{OB2}
\nccurve[angleA=80,angleB=-135]{>-}{LB}{OB2}
\pscurve(0.012,0.9)(-0.12,1.14)(-0.3,0.48)(-0.12,-0.18)(0.012,0.06)
\nccurve[angleA=55,angleB=-148]{->}{OB1}{MR}
\nccurve[angleA=25,angleB=-120]{MR2}{RT}
\pscurve(1.2,1.02)(1.32,1.14)(1.5,0.48)(1.32,-0.18)(1.2,-0.06)
\nccurve[angleA=155,angleB=-55]{ML}{OB2}
\nccurve[angleA=155,angleB=-55]{>-}{ML}{OB2}
\nccurve[angleA=120,angleB=-25]{RB}{ML}
\rput[bl](-0.18,0.57){$\overline{b}$}
\rput[bl](0.24,0.84){$b$}
\rput[bl](-0.18,0.09){$\overline{b}$}
\rput[bl](0.24,-0.09){$b$}
\rput[bl](0.45,1.14){\rnode{A1}{$a$}}
\rput[bl](0.93,1.14){\rnode{A2}{$\overline{a}$}}
\ncarc[nodesepB=2pt]{->}{OA1}{A1}
\ncarc[nodesepB=1.5pt]{OA1}{A1}
\nccurve[angleA=25,angleB=-120,border=1.5pt]{MR}{RT}
\ncarc[nodesepA=2pt,linewidth=0,border=2.5pt]{A2}{OA1}
\ncarc[nodesepA=2pt]{<-}{A2}{OA1}
\ncarc[nodesepA=1.5pt]{A2}{OA1}
\pnode(0.684,0.72){WW}
\ncarc{OA1}{WW}
\rput[bl](0.45,-0.42){\rnode{A3}{$a'$}}
\rput[bl](0.93,-0.42){\rnode{A4}{$\overline{a'}$}}
\ncarc[nodesepA=2pt,border=2pt]{A3}{OA2}
\ncarc[nodesepB=2pt,border=2pt]{OA2}{A4}
\pnode(0.61,-0.06){XX}
\pnode(0.97,-0.06){YY}
\pnode(0.696,0.24){ZZ}
\ncarc[nodesepA=2pt]{A3}{OA2}
\ncarc{>-}{XX}{OA2}
\nccurve[angleA=120,angleB=-25,border=1.5pt]{RB}{ML}
\ncarc[nodesepB=2pt,border=2pt]{OA2}{A4}
\ncarc{-<}{OA2}{YY}
\ncarc{ZZ}{OA2}
\endpspicture
\right]
\label{eq:density}
\end{multline}%
This result simply reflects the fact that all that matters after averaging over
measurement outcomes is that the probe particle passes between $A$ and
$\overline{A}$ with probability $\left| t_{1}\right| ^{2}$, and passes around them
with probability $\left| r_{1}\right| ^{2}$. Since they are initially
unentangled, each additional probe particle has the same analysis
as the first, and just results in another loop that passes between $A$ and
$\overline{A}$ with probability $\left| t_{1}\right| ^{2}$. Noting that an
unlinked $b$ loop may be replaced by a factor $d_{b}$, we see that the reduced
density matrix for $A$ after passing $N$ probe particles through the
interferometer is%
\begin{widetext}
\begin{subequations}
\begin{eqnarray}
\rho _{N}^{A}&=&\sum\limits_{a,a^{\prime }}\frac{A_{a}A_{a^{\prime }}^{\ast }%
}{\sqrt{d_{a}d_{a^{\prime }}}}\sum\limits_{n=0}^{N}\dbinom{N}{n}\left|
r_{1}\right| ^{2\left( N-n\right) }\left| t_{1}\right| ^{2n}\frac{1}{%
d_{b}^{n}}
\pspicture[0.45](0.1,-0.9)(1.85,0.9)
  \psset{linewidth=0.9pt,linecolor=black,arrowscale=1.5,arrowinset=0.15}
  \psline[linearc=.25](0.15,0.8)(0.3,0.2)(1.7,0.2)(1.85,0.8)
  \psline{>-}(0.25,0.4)(0.15,0.8)
  \psline{>-}(1.75,0.4)(1.85,0.8)
  \psline[linearc=.25](0.15,-0.8)(0.3,-0.2)(1.7,-0.2)(1.85,-0.8)
  \psline{<-}(0.25,-0.4)(0.15,-0.8)
  \psline{<-}(1.75,-0.4)(1.85,-0.8)
  \psline[linearc=.2,border=2pt]{->}(0.45,-0.3)(0.45,-0.5)
	(0.8,-0.5)(0.8,0.1)
  \psline[linearc=.2,border=2pt](0.45,0.3)(0.45,0.5)
	(0.8,0.5)(0.8,0.05)
 \psline[linearc=.2](0.44,0.1)(0.44,-0.1)
  \psline[linearc=.2,border=2pt]{->}(1.2,-0.3)(1.2,-0.5)
	(1.55,-0.5)(1.55,0.1)
  \psline[linearc=.2,border=2pt](1.2,0.3)(1.2,0.5)(1.55,0.5)(1.55,0.05)
 \psline[linearc=.2](1.19,0.1)(1.19,-0.1)
  \rput[bl](0.45,0.5){$\overbrace{\quad\ldots\quad}^{n \times b}$}
  \rput[bl]{0}(0.25,0.8){$a$}
  \rput[br]{0}(1.75,0.8){$\overline{a}$}
  \rput[bl]{0}(0.25,-0.9){$a'$}
  \rput[br]{0}(1.68,-0.9){$\overline{a'}$}
  \endpspicture
\\
&=&\sum\limits_{a,a^{\prime }}\frac{A_{a}A_{a^{\prime }}^{\ast }}{\sqrt{d_{a}d_{a^{\prime }}}}\sum\limits_{\left( e,\alpha ,\beta \right) }\left[
\left( F_{a^{\prime },\overline{a^{\prime }}}^{a,\overline{a}}\right) ^{-1}%
\right] _{1,\left( e,\alpha ,\beta \right) }\sum\limits_{n=0}^{N}\dbinom{N}{n%
}\left| r_{1}\right| ^{2\left( N-n\right) }\left| t_{1}\right| ^{2n}\frac{1}{%
d_{b}^{n}}
\pspicture[0.45](0.0,-0.9)(2.2,0.9)
  \psset{linewidth=0.9pt,linecolor=black,arrowscale=1.5,arrowinset=0.15}
  \psline(0.15,0.6)(0.35,0.1)
  \psline{>-}(0.275,0.3)(0.15,0.6)
  \psline{>-}(1.765,0.3)(1.85,0.6)
  \psline(0.15,-0.6)(0.35,0.1)
  \psline(1.65,-0.1)(1.85,-0.6)
  \psline{<-}(0.235,-0.3)(0.15,-0.6)
  \psline{<-}(1.725,-0.3)(1.85,-0.6)
  \psline(0.9,0.0155)(1.1,-0.01523)
  \psline(1.6,-0.09215)(1.65,-0.1)
  \psline[linearc=.2,border=2pt]{->}(0.45,-0.0)(0.45,-0.3)
	(0.8,-0.3)(0.8,0.1)
  \psline[linearc=.2,border=2pt](0.45,0.1)(0.45,0.3)
	(0.8,0.3)(0.8,0.05)
  \psline(0.35,0.1)(0.41,0.0907692)
  \psline[border=2pt](0.41,0.0907692)(0.68,0.0492306)
  \psline(1.65,-0.1)(1.85,0.6)
  \psline[linearc=.2,border=2pt]{->}(1.15,-0.16)(1.15,-0.3)
	(1.5,-0.3)(1.5,0.1)
  \psline[border=2pt](0.9,0.0155)(1.38,-0.0583)
  \psline{<-}(1.05,-0.00754)(1.38,-0.0583)
  \psline[linearc=.2,border=2pt](1.15,0.1)(1.15,0.3)(1.5,0.3)(1.5,0.05)
  \rput[bl](0.45,0.33){$\overbrace{\;\;\;\,\ldots\;\;\;}^{n \times b}$}
  \rput[bl]{0}(0.25,0.6){$a$}
  \rput[br]{0}(1.75,0.6){$\overline{a}$}
  \rput[bl]{0}(0.25,-0.7){$a'$}
  \rput[br]{0}(1.68,-0.7){$\overline{a'}$}
  \rput[br]{0}(1.07,-0.25){$e$}
  \rput[br]{0}(0.25,0){$\alpha$}
  \rput[br]{0}(1.97,-0.25){$\beta$}
  \endpspicture
\\
&=&\sum\limits_{a,a^{\prime }}\frac{A_{a}A_{a^{\prime }}^{\ast }}{\sqrt{d_{a}d_{a^{\prime }}}}\sum\limits_{\left( e,\alpha ,\beta \right) ,\left(
f,\mu ,\nu \right) }\left[ \left( F_{a^{\prime },\overline{a^{\prime }}}^{a,%
\overline{a}}\right) ^{-1}\right] _{1,\left( e,\alpha ,\beta \right) }\left(
\left| r_{1}\right| ^{2}+\left| t_{1}\right| ^{2}M_{be}\right) ^{N}\left[
F_{a^{\prime },\overline{a^{\prime }}}^{a,\overline{a}}\right] _{\left(
e,\alpha ,\beta \right) ,\left( f,\mu ,\nu \right) }
\pspicture[0.43](-0.05,-0.4)(1.5,1)
  \psset{linewidth=0.9pt,linecolor=black,arrowscale=1.5,arrowinset=0.15}
  \psline{->}(0.7,0)(0.7,0.45)
  \psline(0.7,0)(0.7,0.55)
  \psline(0.7,0.55) (0.25,1)
  \psline{->}(0.7,0.55)(0.3,0.95)
  \psline(0.7,0.55) (1.15,1)	
  \psline{->}(0.7,0.55)(1.1,0.95)
  \rput[bl]{0}(0.38,0.2){$f$}
  \rput[br]{0}(1.45,0.8){$\overline{a}$}
  \rput[bl]{0}(0,0.8){$a$}
  \rput[bl]{0}(0.85,0.35){$\mu$}
  \psline(0.7,0) (0.25,-0.45)
  \psline{-<}(0.7,0)(0.35,-0.35)
  \psline(0.7,0) (1.15,-0.45)	
  \psline{-<}(0.7,0)(1.05,-0.35)
  \rput[br]{0}(1.5,-0.4){$\overline{a^{\prime}}$}
  \rput[bl]{0}(-0.05,-0.4){$a^{\prime}$}
  \rput[bl]{0}(0.85,-0.05){$\nu$}
  \endpspicture
  \label{eq:rhoAN}
\end{eqnarray}
\end{subequations}
\end{widetext}%
where the relations in Eq.~(\ref{eq:loopaway}) were used to remove all the $b$
loops, allowing us to perform the sum over $n$, before applying $F$ in the last
step. The intermediate charge label $e$ represents the difference between the
charges $a$ and $a^{\prime }$, taking values that may be fused with
$a^{\prime} $ to give $a$ (the $F$-symbols vanish otherwise). Notice the
potential for this process to transfer an overall anyonic charge $f$ to the
$A$--$\overline{A}$ system.

{}From this result we see, noting $\left| t_{1}\right| ^{2}+\left|
r_{1}\right| ^{2}=1$, that taking the limit $N\rightarrow \infty $ will
exponentially kill off the $e$-channels with $M_{be}\neq 1$, and preserve only
those which have trivial monodromy with $b$, $M_{be}=1$. The interpretation of
$M_{be}=1$ is that $a$ and $a^{\prime }$ have a difference charge $e$ that is
invisible (in the sense of monodromy) to the charge $b$, and so the
corresponding fusion channel remains unaffected by the probe. In general, the
only $e$-channels guaranteed to always survive this process (even for the most
general $B_{k}$ states) have trivial monodromy with all charges. This always
includes $e=1$ (and for modular theories/TQFTs only includes $e=1$), which
requires that $a=a^{\prime }$. Tracing over the $A$ and $\overline{A}$ particles
gives $Tr\left[ \rho _{N}^{A}\right] =1$ as expected, but by considering the
intermediate channels, one also finds that the entire contribution to this trace
is from $e=1$. We should also note that some terms may alternatively be killed
off due to their corresponding $F$-symbols having zero values.

Defining $\rho ^{A}\equiv \lim_{N\rightarrow \infty }\rho _{N}^{A}$, and
denoting by $e_{b}$ the intermediate charges that have trivial monodromy with
$b$, we get the final result (converted back into bra/ket notation, with an
extra factor of $d_f$ inserted for compatibility with the ordinary trace)
\begin{eqnarray}
\rho ^{A} =\sum\limits_{a,a^{\prime }}A_{a}A_{a^{\prime }}^{\ast
}\sum\limits_{\left( e_{b},\alpha ,\beta \right) ,\left( f,\mu ,\nu \right) }
\left[ \left( F_{a^{\prime },\overline{a^{\prime }}}^{a,\overline{a}}\right)
^{-1}\right] _{1,\left( e_{b},\alpha ,\beta \right) }
\nonumber
\\
\times \left[ F_{a^{\prime },\overline{a^{\prime }}}^{a,\overline{a}}\right]
_{\left( e_{b},\alpha ,\beta \right) ,\left( f,\mu ,\nu \right) } \sqrt{d_{f}}
\left| a,\overline{a};f,\mu \vphantom{\overline{a^{\prime }}}\right\rangle
\left\langle a^{\prime },\overline{a^{\prime }};f,\nu \right|\!.
\end{eqnarray}

We now return to the case of general probe particle states as given in Eq.
(\ref{eq:stateBk}). Since tracing requires the charge on a line to match up, a
similar analysis as before applies. For the result, we simply replace $\left(
\left| r_{1}\right| ^{2}+\left| t_{1}\right| ^{2}M_{be}\right) ^{N}$ in
Eq. (\ref{eq:rhoAN}), with%
\begin{equation}
\prod\limits_{k=1}^{N}\left[ 1-\sum\limits_{b}\left| B_{b,\rightarrow
}^{\left( k\right) }t_{1}+B_{b,\uparrow }^{\left( k\right) }r_{1}^{\ast
}\right| ^{2}\left( 1-M_{be}\right) \right] .
\end{equation}%
This term determines the rate at which the $A$ system decoheres, and will
generically vanish exponentially as $N\rightarrow \infty $ unless $e$ has
trivial monodromy (in which case this term simply equals $1$). In some cases,
complete decoherence may even be achieved with a single probe step. By
setting $\left| r_{1}\right| = 0$ and $\left| t_{1}\right| = 1$ in Eq.
(\ref{eq:rhoAN}), we may do away with the interferometer and interpret the
result as decoherence from stray anyons passing between $A$ and
$\overline{A}$, which is important to consider as a source of errors in a
quantum computation.

As a practical example, we apply the results to the Ising anyon model (see e.g. Table 1 of \cite{Kitaev06a} for details), which captures the essence of the Moore-Read state's non-Abelian statistics. For the initial state $\left|
\Psi _{0}\right\rangle =A_{1}\left| 1,1;1\right\rangle +A_{\psi }\left| \psi
,\psi ;1\right\rangle +A_{\sigma }\left| \sigma ,\sigma ;1\right\rangle $, using $b=\sigma $ probes (which have trivial monodromy only with $e=1$) gives%
\begin{eqnarray}
\rho ^{A} = \left| A_{1}\right| ^{2}\left| 1,1;1\right\rangle \left\langle
1,1;1\right| +\left| A_{\psi }\right| ^{2}\left| \psi ,\psi ;1\right\rangle
\left\langle \psi ,\psi ;1\right|
\nonumber
\\
+\left| A_{\sigma }\right| ^{2}\frac{1}{2}\left( \left| \sigma ,\sigma
;1\right\rangle \left\langle \sigma ,\sigma ;1\right| +\left| \sigma ,\sigma
;\psi \right\rangle \left\langle \sigma ,\sigma ;\psi \right| \right)
\end{eqnarray}%
which exhibits loss of all coherence. For $b=\psi $ probes (which have trivial monodromy with both $e=1$ and $\psi$) the result%
\begin{eqnarray}
\rho ^{A} &=&\left| A_{1}\right| ^{2}\left| 1,1;1\right\rangle \left\langle
1,1;1\right| +A_{\psi }A_{1}^{\ast }\left| \psi ,\psi ;1\right\rangle
\left\langle 1,1;1\right|
\nonumber
\\
&&+A_{1}A_{\psi }^{\ast }\left| 1,1;1\right\rangle \left\langle \psi ,\psi
;1\right| +\left| A_{\psi }\right| ^{2}\left| \psi ,\psi ;1\right\rangle
\left\langle \psi ,\psi ;1\right|
\nonumber
\\
&&+\left| A_{\sigma }\right| ^{2}\left| \sigma ,\sigma ;1\right\rangle
\left\langle \sigma ,\sigma ;1\right|
\end{eqnarray}%
shows decoherence only between $\sigma$ and the other charges.

For another example, we consider the Fibonacci anyon model (see e.g.
\cite{Preskill-lectures} for details). The initial
state $\left| \Psi _{0}\right\rangle =A_{1}\left| 1,1;1\right\rangle
+A_{\varepsilon }\left| \varepsilon ,\varepsilon ;1\right\rangle $ probed by
$b=\varepsilon $ particles gives%
\begin{eqnarray}
\rho ^{A} &=&\left| A_{1}\right| ^{2}\left| 1,1;1\right\rangle \left\langle
1,1;1\right| \\
&&+\left| A_{\varepsilon }\right| ^{2}\left( \phi ^{-2}\left| \varepsilon
,\varepsilon ;1\right\rangle \left\langle \varepsilon ,\varepsilon ;1\right|
+\phi ^{-1}\left| \varepsilon ,\varepsilon ;\varepsilon \right\rangle
\left\langle \varepsilon ,\varepsilon ;\varepsilon \right| \right)
\nonumber
\end{eqnarray}%
(where $\phi = \frac{1+\sqrt{5}}{2}$), which exhibits loss of all coherence.

The decoherence effect described in this letter is due to measurements being
made by probe particles. Keeping track of these measurement outcomes, one
generically finds collapse of the target system state into subspaces where the
difference charge has trivial monodromy with the probes \cite{Bonderson07b}.
If this includes only the $e=1$ subspaces, the target
collapses onto a state of definite charge. One may also consider completely
general initial $A$ and $B_{k}$ systems described by density matrices, but as
long as they are all still unentangled, the resulting behavior is
qualitatively similar. It may also be physically relavant in some cases
to allow initial entanglement between the probes, though this greatly
complicates the analysis and results. These generalizations will be
addressed in \cite{Bonderson07b}.

\begin{acknowledgments}
We thank A.~Kitaev, I.~Klich, and especially J.~Preskill for illuminating
discussions, and the organizers and participants of the KITP Workshop on
Topological Phases and Quantum Computation where this work was initiated. We
would also like to acknowledge the hospitality of the IQI, the KITP, and
Microsoft Project Q. This work was supported in part by the NSF under Grant
No.~PHY-0456720 and PHY99-07949, and the NSA under ARO Contract
No.~W911NF-05-1-0294.
\end{acknowledgments}

\bibliographystyle{apsrev}
\bibliography{corr}

\end{document}